\documentclass[a4paper,fleqn,usenatbib]{mnras}

\usepackage{newtxtext,newtxmath}

\usepackage[T1]{fontenc}
\usepackage{ae,aecompl}

\usepackage{graphicx}	
\usepackage{amsmath}	
\usepackage{amssymb}	

\title[Gas mass in cluster core]{The inner gas mass-temperature profile in the core of nearby galaxy clusters}

\author[Haonan Liu et al.]{
Haonan Liu,$^{1}$\thanks{E-mail: hl479@cam.ac.uk}
Andrew C. Fabian$^{1}$
and
Ciro Pinto$^{1,2}$
\\
$^{1}$Institute of Astronomy, Madingley Road, CB3 0HA Cambridge, United Kingdom\\
$^{2}$INAF-IASF Palermo, Via U. La Malfa 153, I-90146 Palermo, Italy\\
}

\date{Accepted XXX. Received YYY; in original form ZZZ}

\pubyear{2020}

\begin{document}
\label{firstpage}
\pagerange{\pageref{firstpage}--\pageref{lastpage}}
\maketitle

\begin{abstract}
We present a mass-temperature profile of gas within the central 10 kpc of a small sample of cool core clusters. 
The mass of the hottest gas phases, at 1.5 and 0.7 keV, is determined from X-ray spectra from the XMM Reflection Grating Spectrometers. 
The masses of the partially ionised atomic and the molecular phases are obtained from published H$\alpha$ and CO measurements.
We find that the mass of gas at 0.7 keV in a cluster is remarkably similar to that of the molecular gas.
Assuming pressure equilibrium between the phases, this means that they occupy volumes differing by $10^5$. 
The molecular gas is located within the H$\alpha$ nebula which is often filamentary and coincides radially and in position angle with the soft X-ray emitting gas.
\end{abstract}

\begin{keywords}
X-rays: galaxies: clusters - galaxies: clusters: general
\end{keywords}



\section{Introduction}

The radiative cooling time is much shorter than a few billion years for the intracluster gas in the core of relaxed galaxy clusters.  
In the absence of a balancing heat source,
the core becomes cooler over time leading to a strong and pressure-driven radiative cooling flow. 
Early analysis of RGS observations showed little evidence of cool (<2 keV) X-ray gas 
(e.g. \citealt{2001A&A...365L..99K}; \citealt{2001A&A...365L.104P}; \citealt{2001A&A...365L..87T}), 
which contradicts the prediction of X-ray cooling rates of up to a few thousand $\rm M_{\odot}\,\rm yr^{-1}$ (e.g. \citealt{1997MNRAS.292..419W}; \citealt{1998MNRAS.298..416P}; \citealt{2001MNRAS.322..589A}). 
Cooling is therefore suppressed by some source of heating, 
with AGN feedback being the most likely source in the cluster core (for reviews on AGN feedback, see \citealt{2012ARA&A..50..455F} and \citealt{2012NJPh...14e5023M}). 
On the other hand, Chandra images show that cooler gas does exist in the central core (e.g. \citealt{2008MNRAS.385.1186S}; \citealt{2010MNRAS.407.2063W}). 
The physical extent of the cooler gas between $\sim 0.5 $ and $\sim 1 $ keV is generally less than 10 kpc radius in many moderately massive clusters (e.g. A3526, the Centaurus cluster, \citealt{2008MNRAS.385.1186S}). 
More extreme cases such as the Perseus cluster are often accompanied by a more extended optical line-emitting nebula (e.g. \citealt{2003MNRAS.344L..48F, 2016MNRAS.461..922F}). 

We have shown that, from a two-temperature model, the cool core has a temperature of around 0.7 keV and the hot ICM temperature is between 1.5 and 3 keV in most clusters (\citealt{2019MNRAS.485.1757L}). 
{O\,\scriptsize{VII}} line emission, characteristic of gas with $kT<0.2$ keV, has been revealed by \citet{2011MNRAS.412L..35S} and \citet{2014A&A...572L...8P,2016MNRAS.461.2077P} in the most central region of clusters with the XMM-\textit{Newton} Reflection Grating Spectrometer (RGS).
The existence of gas at all X-ray temperatures suggests the inexact balance between the cooling and AGN heating. 
The cooling rates deduced from RGS spectra are very low, typically less than 10$\%$ of the predicted rates in the absence of heating (\citealt{2019MNRAS.485.1757L}).
Some cooling is presumably necessary, 
if the central supermassive black hole is powered by gas accretion and a sustaining accretion-feedback loop. 
This would suggest that the multi-phasedness of gas (i.e. the mass-temperature profile) is an important step in completing the understanding of the feedback process.

A complete mass-temperature profile should contain at least four phases: the mass and volume dominating hot ionised gas, a soft X-ray emitting phase, an intermediate phase where the gas is partially ionised and the cold molecular gas. 
The bulk of the hot X-ray emitting ICM extends to over a few hundreds of kpc.
In order to probe any `soft' X-ray gas below the bulk temperature, 
we search for the strongest emission lines that peaks at low temperatures such as {Fe\,\scriptsize{XVII}} and {O\,\scriptsize{VII}} lines. 
These lines require the high spectral resolution of RGS, 
which has a limited spatial resolution in the cross dispersion direction. 
This means it is possible to resolve the core in nearby ($z<0.1$) clusters.

Most cool core clusters are found to have an H$\alpha$ emission nebula. 
The filamentary structure of the nebula spatially coincides with the soft X-ray components 
(e.g. Perseus: \citealt{2003MNRAS.344L..48F,2006MNRAS.366..417F}; Centaurus: \citealt{2005MNRAS.363..216C}, \citealt{2016MNRAS.461..922F}; A1795: \citealt{2001MNRAS.321L..33F}, \citealt{2005MNRAS.361...17C}).
From the spatial coincidence, \citet{2003MNRAS.344L..48F} suggested that the soft X-ray gas is likely mixing with the cold nebula
and the interpenetration of the hot and cold gas leads to the creation of fast particles in the cold gas (\citealt{2011MNRAS.417..172F}). 
The fast particles can then heat and excite the cold gas, powering the observed nebulosity (\citealt{2009MNRAS.392.1475F}). 
In the far UV, the {O\,\scriptsize{VI}} line emission indicates the presence of $10^{5.5}-10^{6}$ K gas, just below the minimum temperatures in X-rays. 
It has been observed in just a few clusters, e.g. the Perseus cluster (\citealt{2005ApJ...635.1031B}), A1795 (\citealt{2005ApJ...635.1031B}; \citealt{2014ApJ...791L..30M}) and the Phoenix cluster (\citealt{2015ApJ...811..111M, 2019ApJ...885...63M}). 

At the lowest temperatures, massive $10^8-10^{10}\,\rm M_{\odot}$ reservoirs of cold ($\sim50$ K) molecular gas have been observed through CO lines in many clusters (\citealt{2001MNRAS.328..762E}; \citealt{2003A&A...412..657S, 2011A&A...531A..85S}; \citealt{2019MNRAS.490.3025R}; \citealt{2019A&A...631A..22O}). 
For most moderately cooling clusters with a cooling rate between a few to $\sim 10\,\rm M_{\odot}\rm\,yr^{-1}$, the bulk of the molecular gas is located within a few kpc of the centre.
The molecular gas exists in a filamentary structure which is likely to form \textit{in situ} from X-ray cooling (e.g. \citealt{2014ApJ...785...44M}; \citealt{2019MNRAS.490.3025R}). 
Molecular hydrogen is also found at a higher temperature of $\sim 2000$ K (\citealt{2002MNRAS.337...63W}). 
It is observed from molecular hydrogen lines at around 2 $\mu$m (e.g. \citealt{2000ApJ...545..670D}; \citealt{2005MNRAS.359..755W}; \citealt{2005MNRAS.358..765H}). 
However, it has significantly less mass than the cold molecular gas (\citealt{2005MNRAS.359..755W}; \citealt{2007MNRAS.382.1246J}).

In a nutshell, a 10 kpc radius spherical core contains most of the gas below 1 keV down to the molecular phase, 
and the filaments of each phase are likely entangled. 
In this work, we select 9 nearby cool core clusters from the CHEmical Enrichment RGS Sample (CHEERS, \citealt{2015A&A...575A..38P}; \citealt{2017A&A...607A..98D}), all of which have deep ($>$100 ks) RGS spectra. 
We assume the following cosmological parameters: $H_{0} = 73 \rm \ km^{-1}Mpc^{-1}$, $\Omega_{\rm M} = 0.27$, $\Omega_{\rm \Lambda} = 0.73$. 
Literature values are corrected using the same cosmology.

\section{Data}

The \textit{XMM-Newton} observatory has the X-ray grating spectrometer RGS and the European Photon Imaging Cameras (EPIC) onboard. 
We performed our spectral analysis using the RGS spectra. 
We followed the data reduction procedure used by \citet{2015A&A...575A..38P} with the \textit{XMM-Newton} Science Analysis System v 16.1.0.
The background was subtracted using template background files based on count rates in CCD 9. 
The background-subtracted spectra were stacked through task \textit{rgscombine}, where the products were converted to SPEX usable format via task \textit{trafo}.
Since RGS is a slitless spectrometer, the line emission is broadened by the spatial extent of the source by 
$\Delta\lambda = 0.138\Delta\theta/m$\,\AA, where $\Delta\lambda$ is the wavelength shift, 
$\Delta\theta$ is the angular offset from the central source in arcmin and $m$ is the spectral order. 
Such a broadening effect dominates over the intrinsic velocity for nearby ($z<0.1$) objects (\citealt{2015A&A...575A..38P}), and so needs to be accounted for in our analysis. 
We corrected the total line broadening using the surface brightness profile of the sources from EPIC/MOS 1.

To spatially resolve extended sources in RGS, we draw a long slit in the dispersion direction. 
The width of the slit is tuned by varying the fraction of point spread function (PSF) included in the cross dispersion direction. 
For a core radius of 10 kpc, the slit width must be 20 kpc. 
The slit contains the inner cool core as well as hot ambient ICM. 
To select our sources from the CHEERS sample, we require both the H$\alpha$ and molecular gas measurements available from the literature. 
The main information is listed in Table \ref{tab:1}. 
For the most distant cluster AS1101, the 20 kpc slit translates to an angular width of 0.305 arcmin, which is still safe to be resolved by RGS. 
Additionally, these clusters have at least 10000 counts in the 10 kpc RGS spectrum.  

We used SPEX version 3.05.00 for spectral analysis with its default proto-Solar abundances of \citet{2009M&PSA..72.5154L}. 
Since the narrow 20 kpc RGS slit includes a limited number of counts, we used C-statistics. 
We adopt 1$\sigma$ ($\Delta C=1$) uncertainty for measurements and 2$\sigma$ ($\Delta C=2.71$) for upper/lower limits, unless otherwise stated.
We use the first order spectra and include the 7-28 \AA\, wavelength band. 
To ensure the minimum bin size is the same as the RGS spectral resolution, 
the spectra are overbinned by a factor of 3.
We further set O, Ne and Fe as free parameters, and couple Mg to Ne and other metals to Fe. 
The spectra are modelled by collisional ionisation equilibrium components (\textit{cie}), 
using multiple combinations of \textit{cie} in SPEX which calculates X-ray emission from a collisionally-ionised plasma at a given temperature $T$ and emission meausure $EM=n_{\rm e}\,n_{\rm H}\,V$. 
Each \textit{cie} component is modified by redshift, galactic absorption with a cold temperature of 0.5 eV and solar abundances (\citealt{2013A&A...551A..25P}).
It is then convolved with the line broadening (\textit{lpro}) component in SPEX. 
These settings except binning are the same as \citet{2019MNRAS.485.1757L}.

\begin{table*}
	\centering
	\caption{Targets, observations and known properties.}
	\label{tab:1}
	\begin{tabular}{lccccccccr} 
		\hline
		Name       & Redshift $^{\it a}$& Observations                         & Total clean time (ks)\,$^{\it b}$ & $\rm N_{H,tot}$\,$^{\it c}$ & $\theta=20\rm\,kpc/\it\,D_{A}$ (arcmin)\,$^{\it d}$ & xPSF\,$^{\it e}$\\\hline
		2A0335+096 & 0.0363   & 0109870101/0201 0147800201           & 137              & 30.7 &0.475 & 74   \\
		A262       & 0.0174   & 0109980101/0601 0504780101/0201      & 113              & 7.15 &0.968 & 90.5 \\
		A496       & 0.0329   & 0135120201/0801 0506260301/0401      & 132              & 6.12 &0.522 & 77.5 \\
		A2052      & 0.0355   & 0109920101 0401520301/0501/0601/0801 & 123              & 3.03 &0.485 & 74.5 \\
                   &          & 0401520901/1101/1201/1301/1601/1701  &                  &      &      &      \\
		A3526      & 0.0114   & 0046340101 0406200101                & 159              & 12.2 &1.47  & 94.5 \\
		A3581      & 0.023    & 0205990101 0504780301/0401           & 147              & 5.32 &0.737 & 86.5 \\
		AS1101     & 0.0580   & 0147800101 0123900101                & 100              & 1.17 &0.305 & 56   \\
		Perseus    & 0.0179   & 0085110101/0201 0305780101           & 181              & 20.7 &0.942 & 90   \\
		Virgo      & 0.0043   & 0114120101 0200920101                & 190              & 2.11 &4.06  & 99   \\
		\hline
	\end{tabular}
	\\${(a)}$ The redshifts are taken from the NED database (https://ned.ipac.caltech.edu/).
	  ${(b)}$ RGS net exposure time.
      ${(c)}$ Total hydrogen column densities in $10^{20}\rm\, cm^{-2}$ (see http://www.swift.ac.uk/analysis/nhtot/; \citealt{2005A&A...440..775K}; \citealt{2013MNRAS.431..394W}).
      ${(d)}$ The angular width of a fixed 20 kpc RGS slit region at the redshift of the source.
      ${(e)}$ The percentage of the PSF included in the cross dispersion direction in RGS which corresponds to the angular size in ${(d)}$. 
      \\
\end{table*}

\section{Results \& Discussion}

\subsection{Mass measurement of X-ray components}

It was recently shown that clusters in our sample can be modelled by a two-component model with free temperatures (see e.g. \citealt{2017A&A...607A..98D} and \citealt{2019MNRAS.485.1757L}).
The {Fe\,\scriptsize{XVII}} lines are observable in these clusters, which indicate the temperature of the cooler component is at around 0.7 keV.
On the other hand, using RGS spectroscopy means that a fraction of hot and massive gas is projected along the 20 kpc slit, which can have a higher temperature than the two-component model.
To effectively introduce a third temperature component that accounts for the hotter gas while reducing degeneracy in the emission measure, we fix the temperature of three components at 3, 1.5 and 0.7 keV.
The temperature of these components is separated by at least 0.8 keV and hence they represent distinct gas phases. 
The {O\,\scriptsize{VII}} signature is weak in our sample. 
We find a further component at 0.2 keV usually gives an upper limit. 
Resonant scattering of {Fe\,\scriptsize{XVII}} lines also leads to degeneracy in the emission measure of the 0.2 and 0.7 keV components as found in some clusters in the CHEERS sample (\citealt{2015A&A...575A..38P}; \citealt{2019MNRAS.485.1757L}). 
Therefore we only use the three-component model in this work.

To estimate the gas masses, we assume that the inner core is in hydrostatic equilibrium, and that the different gas phases are in pressure equilibrium with each other at any chosen radii.
We interpolate the radial mean density and temperature profiles from the ACCEPT catalogue (\citealt{2009ApJS..182...12C}) to calculate the pressure at 10 kpc, within which most of the gas at 1.5 and 0.7 keV lies.
The ACCEPT catalogue contains projected temperature profiles.
\citet{2014MNRAS.438.2341P} and \citet{2016MNRAS.460.2625L} have shown that projected temperatures are higher which causes higher core entropy within $\sim 10$ kpc.
This has a similar effect on pressure.
\citet{2017ApJ...837...51H} showed that, in A496, the deprojected temperature at 10 kpc is $\sim 10\%$ lower than the projected temperature while the density is consistent with that reported in the ACCEPT catalogue.
They further showed that the shape of the radial density profile is similar to four other cool core clusters, where the density varies as $1/r^{0.6}$ and the deprojected temperature as $r^{0.3}$. 
This means that, if the radial dependence of temperature and density holds in general in the rest of our sample, the pressure $n_{\rm e}T$ typically rises by a factor of 1.6 to 2.2 at 2 kpc from the value at 10 kpc and is even higher at the innermost 1 kpc.
Exact scaling of radial profiles is often not possible due to the complexity of the inner structure of the core.
Different gas phases are neither spherically symmetric nor completely volume-filling.
For simplicity, we estimate the gas mass using the pressure at 10 kpc in this work.
For a given temperature, the mass of the X-ray emitting component is proportional to the ratio of its emission measure to the pressure.
The net effect is that our masses may be overestimated by a factor of up to 2.

\begin{figure*}
    \includegraphics[width=2\columnwidth]{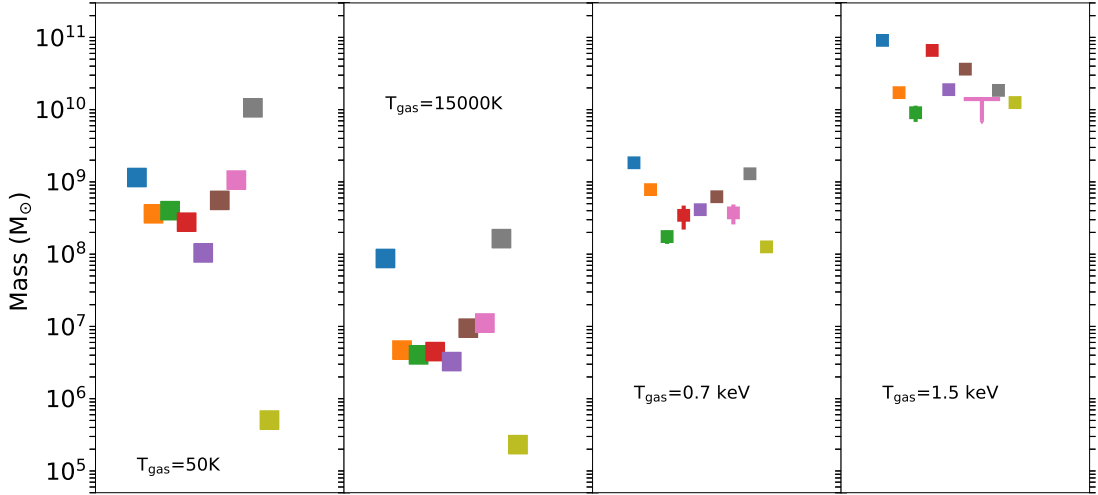}
    \caption{Visualisation of mass-temperature profile presented in Table \ref{tab:1}, 
    with the coolest molecular gas from the left and the hot ambient ICM at 1.5 keV from the right. 
    Within each panel, we shift the data points horizontally for clarity and the clusters are presented in the same order as the first column in Table \ref{tab:1}, i.e. 2A0335+096 is on the left and Virgo is on the right. 
    }
    \label{fig:1}
\end{figure*}

The mass-temperature profile is seen in Fig. \ref{fig:1} (and detailed in Table \ref{tab:2}). 
The RGS slit only selects a fraction of the 3 keV gas, so the mass of the 3 keV component is only a lower limit. 
We see that, for the most relevant components at 1.5 and 0.7 keV, the gas mass decreases by at least an order of magnitude towards lower temperatures.
In 6 out of 9 clusters, the ratio of the mass of these two components is between 20 and 60, with two more with a ratio of 10.
This pattern is likely intrinsically regulated, whereas the reason for the connection is unclear.
The comparison between the gas mass and corresponding emission measure for either the 1.5 or 0.7 keV component also shows that the mass scatter is smaller than that of the emission measure.

The 1.5 and 0.7 keV components do not always have the same spatial extent. 
From Chandra analysis, the 0.7 keV gas exhibits spatial coincidence with the 1.5 keV component in the Centaurus cluster (\citealt{2008MNRAS.385.1186S,2016MNRAS.457...82S}). 
However, the 0.7 keV gas occupies a significantly smaller region in A3581 (\citealt{2005MNRAS.356..237J}), Virgo (\citealt{2010MNRAS.407.2063W}) and Perseus (\citealt{2006MNRAS.366..417F}),
that region mainly lies within the extent of the next hotter component. 
Although the angle-averaged temperature profile of the hot gas drops steadily inward within a cool core, 
the gas is formed of non-spherical phases that occupy complex shapes. 
Some of these shapes may be due to the AGN bubbling process. 
The 0.7 and 1.5 keV components are not volume-filling in a radial sense. 

\citet{2020ApJ...889L...1L} have attempted to measure the spectrum of turbulence in the hot gas of several cool core clusters by using H$\alpha$ velocity measurements of cold gas.
This assumes that the cold gas is a tracer in the hot gas. Our results here, however,\, show that the \textit{total} mass of the cold gas and the hot gas below 1 keV, (typical of the hot gas around the H$\alpha$ filaments) are comparable.
This must surely be taken into consideration in calculations of turbulence and may contribute to the steepness of the velocity function at small radii.

\begin{table*}
	\centering
	\caption{Mass of different temperature components}
	\label{tab:2}
	\begin{tabular}{lcccccccr} 
		\hline
		Name       & $M_{\rm 500}$             & $M_{\rm mol}$              & $M_{\rm H\alpha}$  & $M_{\rm X,0.7\,keV}$ & $M_{\rm X,1.5\,keV}$ & $M_{\rm X,3.0\,keV}$ & Reference\\\hline
		2A0335+096 & 3.45$\times10^{14}$       & 1.2$\pm$0.2$\times10^{9}$  & 8.71$\times10^{7}$ &1.8$\pm$0.2$\times10^{9}$  &9.1$\pm$0.1$\times10^{10}$  &>1.5$\times10^{9}$ & [1,2,5]\\
		A262       & 1.19$\times10^{14}$       & 3.62$\pm$0.02$\times10^{8}$& 4.71$\times10^{6}$ &7.8$\pm$0.8$\times10^{8}$  &1.7$\pm$0.3$\times10^{10}$  &>3.2$\times10^{10}$& [1,3,6]\\
		A496       & 2.91$\times10^{14}$       & 4.0$\pm$0.9$\times10^{8}$  & 4.04$\times10^{6}$ &1.8$\pm$0.4$\times10^{8}$  &9$\pm$2$\times10^{9}$       &>6.4$\times10^{10}$& [3,7]  \\
		A2052      & 2.49$\times10^{14}$       & 2.8$\pm$0.4$\times10^{8}$  & 4.49$\times10^{6}$ &3$\pm$1$\times10^{8}$      &6.6$\pm$0.6$\times10^{10}$  &>1.7$\times10^{10}$& [1,7]  \\
		A3526      & 1.62$\times10^{14}$       & 1.04$\times10^{8}$         & 3.26$\times10^{6}$ &4.0$\pm$0.2$\times10^{8}$  &1.9$\pm$0.1$\times10^{10}$  &>6.3$\times10^{9}$ & [2,8]  \\
		A3581      & 1.08$\times10^{14}$       & 5.55$\times10^{8}$         & 9.46$\times10^{6}$ &6.2$\pm$0.5$\times10^{8}$  &3.6$\pm$0.1$\times10^{10}$  &>4.5$\times10^{8}$ & [2,7]  \\
		AS1101     & 2.83$\times10^{14}$       & 1.06$\times10^{9}$         & 1.12$\times10^{7}$ &4$\pm$1$\times10^{8}$      &<1.4$\times10^{10}$         &>1.6$\times10^{11}$& [2,9]  \\
		Perseus    & 6.15$\times10^{14}$       & 1.06$\times10^{10}$        & 1.64$\times10^{8}$ &1.30$\pm$0.05$\times10^{9}$&1.8$\pm$0.1$\times10^{10}$  &>1.2$\times10^{11}$ & [4,10]\\
		Virgo      & N/A                       & 5.06$\times10^{5}$         & 2.31$\times10^{5}$ &1.26$\pm$0.03$\times10^{8}$&1.26$\pm$0.01$\times10^{10}$&>8.2$\times10^{6}$ & [2,10]  \\
		\hline
	\end{tabular}
	\\The masses are in $\rm M_{\odot}$. $M_{\rm 500}$ are taken from the MCXC catalog (\citealt{2011A&A...534A.109P}). The molecular masses are taken from [1] \citet{2019MNRAS.490.3025R} [2] \citet{2019A&A...631A..22O} [3] \citet{2003A&A...412..657S} [4] \citet{2011A&A...531A..85S}. 
	The references for H$\alpha$ emission are [5] \citet{2007AJ....134...14D} [6] \citet{1999MNRAS.306..857C} [7] \citet{2016MNRAS.460.1758H} [8] \citet{2005MNRAS.363..216C} [9] \citet{2005MNRAS.360..748J} [10] \citet{1989ApJ...338...48H}. 
	 
\end{table*}

\subsection{Core cooling times}

The radiative cooling time at 10 kpc is calculated using the definition 
\begin{equation}
    \label{eqn:1}
    t_{\rm cool}=\frac{3 (n_{\rm e}+n_{\rm i})k_{\rm B}T}{2n_{\rm e}n_{\rm i}\Lambda},
\end{equation}
where $n_{\rm e}$ and $n_i$ are electron and ion densities in $\rm cm^{-3}$ and $\Lambda$ is the cooling function. 
We assume the cooling process is isochoric, and hence 3/2 is required. 
For a fully ionised plasma, $n_H$ is effectively 0.857 times the electron density if we assume the hydrogen mass fraction is 75$\%$. 
The cooling function is determined for the metallicity from the best fit model in SPEX. 
It is the ratio of the total luminosity (we calculate between 0.001 and 1000 keV) to the emission measure. 

The cooling times are listed in Table \ref{tab:3} for gas at 0.7 and 1.5 keV. 
We can compare our results at 1.5 keV with the cooling time at 10 kpc in \citet{2009ApJS..182...12C}. 
Our values are within 40$\%$ lower in 5 clusters and more than a factor of 2 lower in the other half of the sample. 
The difference is mainly due to the ICM temperature being higher than 1.5 keV. 
Ihe cooling time can be 2 times longer for a gas temperature of 2 keV. 
This temperature is commonly seen in the ICM of clusters at 10 kpc in \citet{2009ApJS..182...12C}. 
The radial cooling time profile shows that it can drop by a factor of a few at 2 kpc albeit at a slightly lower temperature. 
Hence, the true cooling time at the fixed temperatures might be lower if the gas is located further in than 10 kpc. 
Notice that the interpolation of the pressure is the major source of uncertainty in the cooling time in Perseus and A2052. 
It is difficult to estimate the density at 10 kpc due to the presence of X-ray cavities (e.g. the Perseus cluster, \citealt{2003MNRAS.344L..43F}).

We can estimate the lifetime of the cool gas at 0.7 keV assuming it is cooling at the rate of the residual cooling flow $\dot M_{\rm C, \,1 cie + 2 cf}$.  
We find that most of these components are long-lived (Table \ref{tab:3}). 
The lifetime are of course longer if AGN heating acts on these components.
Note that in the Perseus cluster, we measured a higher cooling rate from a larger 99$\%$ PSF core spectrum (\citealt{2019MNRAS.485.1757L}).

\begin{table*}
	\centering
	\caption{Pressure, cooling time and residual cooling rate of X-ray components}
	\label{tab:3}
	\begin{tabular}{lcccccccr} 
		\hline
		Name       & $n_{\rm e}T$   & $t_{\rm cool,0.7\,keV}$   & $t_{\rm cool,1.5\,keV}$  & $\dot M_{\rm C, \,1 cie + 2 cf}$ & $\tau_{\rm x}$  \\\hline
		2A0335+096 & 0.12  & 42  & 230 & 12$\pm4$    & 0.15 \\
		A262       & 0.027 & 140 & 910 & 0.9$\pm0.3$ & 0.9  \\
		A496       & 0.097 & 34  & 230 & $<$1.92     & $>$0.09\\
		A2052      & 0.050 & 99  & 550 & $<$1.32     & $>$0.23\\
		A3526      & 0.050 & 57  & 420 & 0.6$\pm0.1$ & 0.7\\
		A3581      & 0.032 & 120 & 770 & 2.1$\pm0.9$ & 0.29\\
		AS1101     & 0.098 & 35  & 240 & $<$5.09     & $>$0.08\\
		Perseus    & 0.18  & 25  & 148 & 31$\pm2$    & 0.04 \\
		Virgo      & 0.042 & 88  & 570 & $<$0.07     & $>$1.8\\
		\hline
	\end{tabular} 
	\\The $n_{\rm e}T$ pressure at 10 kpc is in $\rm keV cm^{-3}$ and the cooling time is in Myr. $\dot M_{\rm C, \,1 cie + 2 cf}$ is the residual cooling rate in $\rm M_{\odot}\,\rm yr^{-1}$ measured between 0.7 and 0.01 keV from \citet{2019MNRAS.485.1757L}. $\tau_{\rm x}$ is the lifetime in Gyr of the X-ray component at 0.7 keV cooling at the rate of $\dot M_{\rm C, \,1 cie + 2 cf}$.   
\end{table*}

\begin{figure*}
    \includegraphics[width=2\columnwidth]{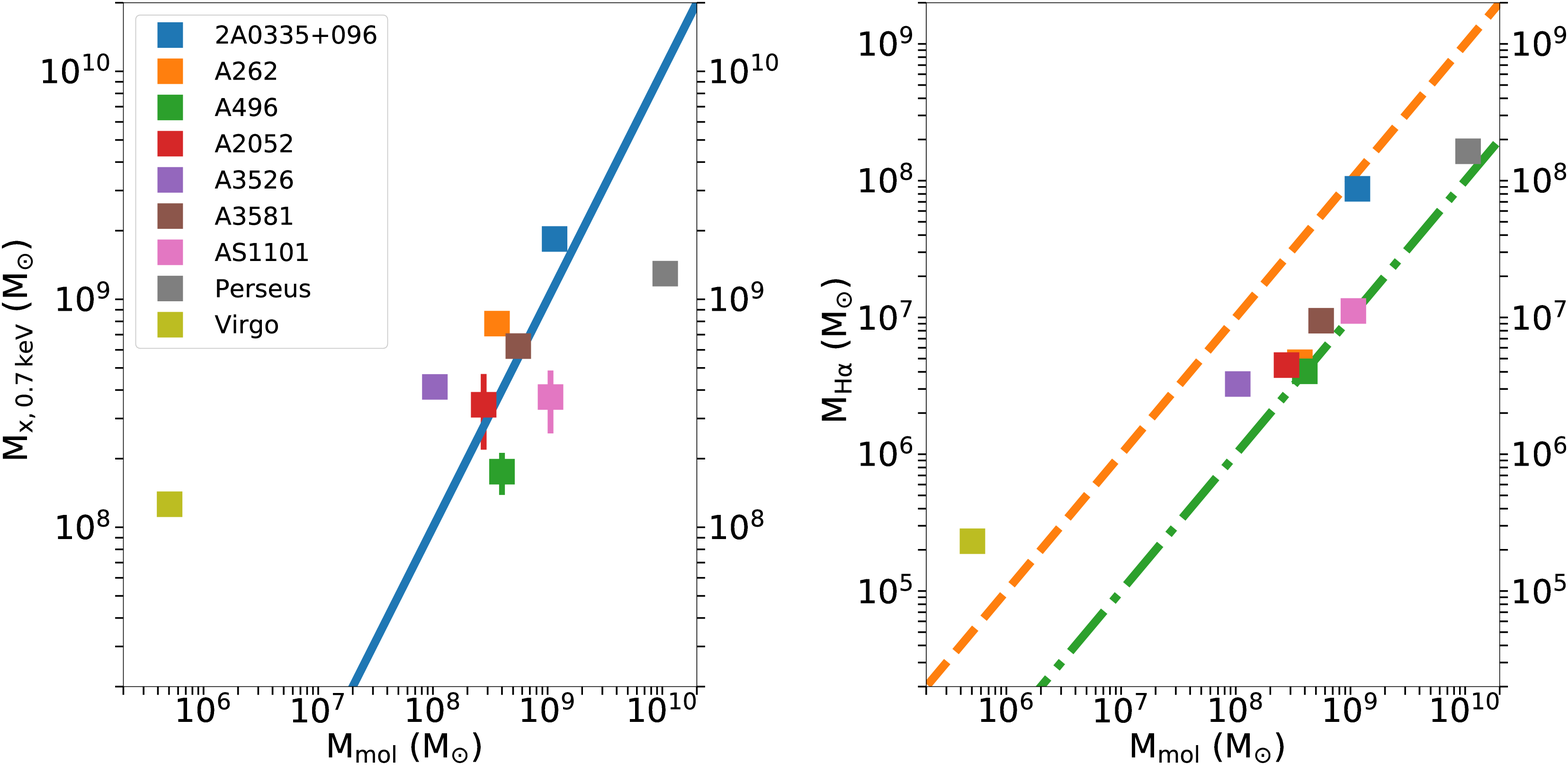}
    \caption{Left: The X-ray emitting gas against molecular mass. 
    The blue line indicates where the molecular mass would match the soft X-ray mass. 
    Right: The mass of H$\alpha$ component against molecular mass. The orange dash line has a constant $M_{\rm H \alpha}/M_{\rm mol}$ ratio of 0.1 and the green dash-dot line has a ratio of 0.01. 
    The uncertainties in the mass of molecular and H$\alpha$ emitting gas is smaller than the marker size. 
    }
    \label{fig:2}
\end{figure*}

\subsection{Mass of optical line-emitting ionised gas}

The optical line-emitting gas is of interest because of the spatial coincidence with the soft X-ray emitting gas at around 0.7 keV (\citealt{2003MNRAS.344L..48F}; see Appendix for details). 
The gas is partially ionised in this phase, which can be traced by H$\alpha$ emission. 
Here, we estimate the approximate gas mass from the observed H$\alpha$ luminosity $L_{\rm H \alpha}$. 
The mass of the H$\alpha$ filaments is 
\begin{equation}
    M_{\rm H \alpha}=\frac{\mu_e\,m_p\,n_e\,L_{\rm H \alpha}}{j_{\rm H \alpha}},
\end{equation}
\noindent where $\mu_e\,m_p$ is the mean mass per electron, $n_e$ is the electron number density.
$j_{\rm H \alpha}$ is the emissivity of the H$\alpha$ line, and is calculated from 
\begin{equation}
    j_{\rm H \alpha}=1.3\times\,10^{-23}\,n_{\rm e}\,n_{\rm H}\,T^{-1/2}\,log(I_H/kT) \rm\,erg\,s^{-1}\, cm^{-3},
\end{equation}
where T is the temperature in K and $I_H$ is the ionisation energy of hydrogen. 
The emissivity of the H$\alpha$ line varies strongly with temperature. 
We assume the H$\alpha$ filaments have a temperature of 15000 K estimated from its peak emissivity. 
We use the pressure at 10 kpc in general. 
In the Virgo and Centaurus clusters, the H$\alpha$ nebula is smaller than 10 kpc in radius (\citealt{2010MNRAS.407.2063W}; \citealt{2005MNRAS.363..216C}; \citealt{2014MNRAS.437..862H}). 
We hence use the pressure at 1 kpc for Virgo, and 5 kpc for Centaurus. 
Note that in A262, the H$\alpha$ luminosity is taken from \citet{1999MNRAS.306..857C} which used long slits.  
The true luminosity can therefore be higher if the filaments of the clusters are more extended than the slit. 

We find that the optical line-emitting gas is two orders of magnitude less massive than the X-ray components in most clusters. 
This mass estimate is consistent with the literature e.g. Virgo (\citealt{1993ApJ...413..531S}). 
Note that since the Perseus cluster has a highly extended H$\alpha$ nebula (\citealt{2001AJ....122.2281C}; \citealt{2008Natur.454..968F}), the average density of the filaments might be lower than the value at 10 kpc. 
This suggests that the H$\alpha$ nebula is likely more massive than our estimate in Perseus. 
{O\,\scriptsize{VI}}, which emits in the FUV, is generally not available for our sample. 
Our estimate of gas masses ignores photoelectric absorption by cooler gas and reddening by dust. 
This will be difficult to determine until we have much detailed spatial information.

\subsection{Connection between soft X-ray and molecular gas}

The most important finding of this work is the comparison between the mass of the soft X-ray gas and the molecular gas. 
The recent ALMA observations showed that cold ($\sim 50$ K) molecular gas is located within the H$\alpha$ nebula (\citealt{2019A&A...631A..22O}; \citealt{2019MNRAS.490.3025R}). 
It is therefore also within the soft X-ray emitting region albeit more compact. 
The molecular gas mass is calculated from the integrated CO intensity by adopting the Galactic value of CO-to-H$_{2}$ ($X_{\rm CO}$) conversion factor and typical CO line ratios in Brightest Cluster Galaxies. 
Since CO is optically thick to radiation, there is an uncertainty in the conversion factor. 
From a $^{13}$CO detection in RXJ0821+0752, \citet{2017ApJ...848..101V} suggested that using the Galactic value for 
$X_{\rm CO}$ likely overestimates the molecular mass by a factor of 2. 
Note that ALMA does not detect faint outer molecular filaments which only have a small fraction of the mass. 
A spread of molecular mass is seen in our sample. 
The molecular gas in the Perseus cluster is 4 orders of magnitude more massive than that of the Virgo cluster.

The comparison between the 0.7 keV and molecular gas mass is seen in Fig. \ref{fig:2}. 
We show that the soft X-ray and molecular gas have similar masses in most clusters. 
This suggests that these two gas phases are strongly linked. 
However, it is not clear to us why this should be the case. 
As we have seen, if mass exchange from the soft X-ray emitting gas is responsible, it takes more than $\sim$10$^8$ yr by radiative cooling in most clusters.
From spatially-resolved images, soft X-ray gas has a much larger extent than the molecular phase which is often only a few arcsec across (e.g. \citealt{2019MNRAS.490.3025R}). 
At the same time, assuming both phases are in pressure equilibrium, the volume of the gas would be proportional to its temperature. 
It requires that the soft X-ray component occupies $10^5$ times larger volume. 
Therefore the connection between these gas phases is not simple. 

The Virgo and Perseus clusters are exceptions in our sample. 
For Virgo, the 0.7 keV gas mass is 250 times higher than the molecular mass. 
It has been suggested the molecular gas could have been destroyed or excited by X-ray shocks, or interacted with the radio lobe (\citealt{2018MNRAS.475.3004S}). 
The deposition rate of the molecular gas can also be reduced by star formation, since the star formation rate is comparable to the X-ray cooling rate (\citealt{2019MNRAS.485.1757L}). 
For the Perseus cluster, the molecular mass is 8 times more massive than 0.7 keV gas. 
Given the large systematic uncertainty in the molecular mass measurements, 
the difference in the mass of the two phases is likely only a factor of 4. 
It can in part be explained by the fact that the molecular filaments are more extended than the 10 kpc core (\citealt{2011A&A...531A..85S}). 

Since we assume gas of different phases are in pressure equilibrium, 
it is important to consider the physical scale of different phases if they have similar masses. 
We use the example of a typical cool core cluster A3581 with moderate X-ray emitting cool gas of $\sim\,10^9\rm M_{\odot}$. 
Pressure equilibrium suggests that the volume ratio of the gas will be the same as the temperature ratio. 
It means that the molecular gas occupies about 5 orders of magnitude smaller volume than the 0.7 keV gas. 
If the mass is put in a spherical ball with a consistent density, we can calculate the radius of that sphere. 
We list the details in Table \ref{tab:4}.
The size of the sphere of the molecular gas is less than 100 pc, even though it extends over 5 kpc (\citealt{2019A&A...631A..22O}). 
The atomic and molecular gas is observed to be in many thin filaments at the highest resolution (e.g. HST image from \citealt{2008Natur.454..968F,2016MNRAS.461..922F}). 
The spatial coincidence of different gas phases hence suggests they are intermingled. 
A magnetic field is required to give integrity to the filaments (e.g \citealt{2016MNRAS.461..922F}), with pressure equilibrium implying $B\sim$ 50$\mu$G.
Note that the resolved HST filaments probably contain many much smaller threads (\citealt{2016MNRAS.461..922F}).

\begin{table}
\caption{Scale of different gas phase in A3581.}  
\label{tab:4}
\begin{tabular}{c c c c c c}     
\hline
Phase      & Mass ($\rm M_{\odot}$) & $T$ (K)   & $n_{\rm e}$ ($\rm cm^{-3}$) \\\hline
Soft X-ray & 6.2        $\times10^{8}$ & 8.2$\times10^{6}$ & 0.046     \\
H$\alpha$  & 9.46       $\times10^{6}$ & 1.5$\times10^{4}$ & 25        \\
Molecular  & 5.55       $\times10^{8}$ & 50                & 7500      \\
\hline       
\end{tabular}
\end{table}

\section{Conclusions}

In this work, we studied the mass-temperature profile of the gas in the inner regions of nearby cool core clusters of galaxies inspired by the similar morphology of the soft X-ray emitting gas, the optical nebula as well as the molecular gas. 
For the X-ray components, we used a three-temperature model to describe the XMM-\textit{Newton} RGS spectra of the core of 9 nearby clusters. 
The gas mass is calculated from the emission measure by assuming pressure equilibrium at the selected radii. 
We showed that the X-ray mass-temperature profile is similar among these clusters particularly in the 1.5 and 0.7 keV components. 
The total X-ray mass below 1 keV is found to be comparable to that of the molecular gas in 7 out of 9 clusters. 
Although the H$\alpha$ nebula is more closely aligned with the soft X-ray components, it only has 0.1-1$\%$ of the mass of the other phases. 
Future models of cluster cores should consider using these relations. 
The molecular gas occupies by far the smallest volume in the form of thin magnetized filaments. 

\section*{Acknowledgements}

This work is based on observations obtained with XMM-\textit{Newton},
an ESA science mission funded by ESA Member States and USA (NASA). 
We thank the referee Brian McNamara for his helpful comments.

\section*{Data availability}

No new data were generated or analysed in support of this research.




\bibliographystyle{mnras}
\bibliography{Haonan_Liu_paper_cluster_mass_temperature} 

\begin{thebibliography}{}
\makeatletter
\relax
\def\mn@urlcharsother{\let\do\@makeother \do\$\do\&\do\#\do\^\do\_\do\%\do\~}
\def\mn@doi{\begingroup\mn@urlcharsother \@ifnextchar [ {\mn@doi@}
  {\mn@doi@[]}}
\def\mn@doi@[#1]#2{\def\@tempa{#1}\ifx\@tempa\@empty \href
  {http://dx.doi.org/#2} {doi:#2}\else \href {http://dx.doi.org/#2} {#1}\fi
  \endgroup}
\def\mn@eprint#1#2{\mn@eprint@#1:#2::\@nil}
\def\mn@eprint@arXiv#1{\href {http://arxiv.org/abs/#1} {{\tt arXiv:#1}}}
\def\mn@eprint@dblp#1{\href {http://dblp.uni-trier.de/rec/bibtex/#1.xml}
  {dblp:#1}}
\def\mn@eprint@#1:#2:#3:#4\@nil{\def\@tempa {#1}\def\@tempb {#2}\def\@tempc
  {#3}\ifx \@tempc \@empty \let \@tempc \@tempb \let \@tempb \@tempa \fi \ifx
  \@tempb \@empty \def\@tempb {arXiv}\fi \@ifundefined
  {mn@eprint@\@tempb}{\@tempb:\@tempc}{\expandafter \expandafter \csname
  mn@eprint@\@tempb\endcsname \expandafter{\@tempc}}}

\bibitem[\protect\citeauthoryear{{Allen}, {Fabian}, {Johnstone}, {Arnaud}  \&
  {Nulsen}}{{Allen} et~al.}{2001}]{2001MNRAS.322..589A}
{Allen} S.~W.,  {Fabian} A.~C.,  {Johnstone} R.~M.,  {Arnaud} K.~A.,   {Nulsen}
  P.~E.~J.,  2001, \mn@doi [\mnras] {10.1046/j.1365-8711.2001.04135.x}, \href
  {http://adsabs.harvard.edu/abs/2001MNRAS.322..589A} {322, 589}

\bibitem[\protect\citeauthoryear{{Blanton}, {Randall}, {Clarke}, {Sarazin},
  {McNamara}, {Douglass}  \& {McDonald}}{{Blanton}
  et~al.}{2011}]{2011ApJ...737...99B}
{Blanton} E.~L.,  {Randall} S.~W.,  {Clarke} T.~E.,  {Sarazin} C.~L.,
  {McNamara} B.~R.,  {Douglass} E.~M.,   {McDonald} M.,  2011, \mn@doi [\apj]
  {10.1088/0004-637X/737/2/99}, \href
  {https://ui.adsabs.harvard.edu/abs/2011ApJ...737...99B} {737, 99}

\bibitem[\protect\citeauthoryear{{Bregman}, {Miller}, {Athey}  \&
  {Irwin}}{{Bregman} et~al.}{2005}]{2005ApJ...635.1031B}
{Bregman} J.~N.,  {Miller} E.~D.,  {Athey} A.~E.,   {Irwin} J.~A.,  2005,
  \mn@doi [\apj] {10.1086/497421}, \href
  {http://adsabs.harvard.edu/abs/2005ApJ...635.1031B} {635, 1031}

\bibitem[\protect\citeauthoryear{{Canning} et~al.,}{{Canning}
  et~al.}{2013}]{2013MNRAS.435.1108C}
{Canning} R.~E.~A.,  et~al., 2013, \mn@doi [\mnras] {10.1093/mnras/stt1345},
  \href {https://ui.adsabs.harvard.edu/abs/2013MNRAS.435.1108C} {435, 1108}

\bibitem[\protect\citeauthoryear{{Cavagnolo}, {Donahue}, {Voit}  \&
  {Sun}}{{Cavagnolo} et~al.}{2009}]{2009ApJS..182...12C}
{Cavagnolo} K.~W.,  {Donahue} M.,  {Voit} G.~M.,   {Sun} M.,  2009, \mn@doi
  [\apjs] {10.1088/0067-0049/182/1/12}, \href
  {https://ui.adsabs.harvard.edu/abs/2009ApJS..182...12C} {182, 12}

\bibitem[\protect\citeauthoryear{{Conselice}, {Gallagher}  \&
  {Wyse}}{{Conselice} et~al.}{2001}]{2001AJ....122.2281C}
{Conselice} C.~J.,  {Gallagher} III J.~S.,   {Wyse} R.~F.~G.,  2001, \mn@doi
  [\aj] {10.1086/323534}, \href
  {http://adsabs.harvard.edu/abs/2001AJ....122.2281C} {122, 2281}

\bibitem[\protect\citeauthoryear{{Crawford}, {Allen}, {Ebeling}, {Edge}  \&
  {Fabian}}{{Crawford} et~al.}{1999}]{1999MNRAS.306..857C}
{Crawford} C.~S.,  {Allen} S.~W.,  {Ebeling} H.,  {Edge} A.~C.,   {Fabian}
  A.~C.,  1999, \mn@doi [\mnras] {10.1046/j.1365-8711.1999.02583.x}, \href
  {http://adsabs.harvard.edu/abs/1999MNRAS.306..857C} {306, 857}

\bibitem[\protect\citeauthoryear{{Crawford}, {Sanders}  \& {Fabian}}{{Crawford}
  et~al.}{2005a}]{2005MNRAS.361...17C}
{Crawford} C.~S.,  {Sanders} J.~S.,   {Fabian} A.~C.,  2005a, \mn@doi [\mnras]
  {10.1111/j.1365-2966.2005.09149.x}, \href
  {http://adsabs.harvard.edu/abs/2005MNRAS.361...17C} {361, 17}

\bibitem[\protect\citeauthoryear{{Crawford}, {Hatch}, {Fabian}  \&
  {Sanders}}{{Crawford} et~al.}{2005b}]{2005MNRAS.363..216C}
{Crawford} C.~S.,  {Hatch} N.~A.,  {Fabian} A.~C.,   {Sanders} J.~S.,  2005b,
  \mn@doi [\mnras] {10.1111/j.1365-2966.2005.09463.x}, \href
  {http://adsabs.harvard.edu/abs/2005MNRAS.363..216C} {363, 216}

\bibitem[\protect\citeauthoryear{{Donahue}, {Mack}, {Voit}, {Sparks}, {Elston}
  \& {Maloney}}{{Donahue} et~al.}{2000}]{2000ApJ...545..670D}
{Donahue} M.,  {Mack} J.,  {Voit} G.~M.,  {Sparks} W.,  {Elston} R.,
  {Maloney} P.~R.,  2000, \mn@doi [\apj] {10.1086/317836}, \href
  {https://ui.adsabs.harvard.edu/abs/2000ApJ...545..670D} {545, 670}

\bibitem[\protect\citeauthoryear{{Donahue}, {Sun}, {O'Dea}, {Voit}  \&
  {Cavagnolo}}{{Donahue} et~al.}{2007}]{2007AJ....134...14D}
{Donahue} M.,  {Sun} M.,  {O'Dea} C.~P.,  {Voit} G.~M.,   {Cavagnolo} K.~W.,
  2007, \mn@doi [\aj] {10.1086/518230}, \href
  {http://adsabs.harvard.edu/abs/2007AJ....134...14D} {134, 14}

\bibitem[\protect\citeauthoryear{{Edge}}{{Edge}}{2001}]{2001MNRAS.328..762E}
{Edge} A.~C.,  2001, \mn@doi [\mnras] {10.1046/j.1365-8711.2001.04802.x}, \href
  {http://adsabs.harvard.edu/abs/2001MNRAS.328..762E} {328, 762}

\bibitem[\protect\citeauthoryear{{Fabian}}{{Fabian}}{2012}]{2012ARA&A..50..455F}
{Fabian} A.~C.,  2012, \mn@doi [\araa] {10.1146/annurev-astro-081811-125521},
  \href {http://adsabs.harvard.edu/abs/2012ARA\%26A..50..455F} {50, 455}

\bibitem[\protect\citeauthoryear{{Fabian}, {Sanders}, {Ettori}, {Taylor},
  {Allen}, {Crawford}, {Iwasawa}  \& {Johnstone}}{{Fabian}
  et~al.}{2001}]{2001MNRAS.321L..33F}
{Fabian} A.~C.,  {Sanders} J.~S.,  {Ettori} S.,  {Taylor} G.~B.,  {Allen}
  S.~W.,  {Crawford} C.~S.,  {Iwasawa} K.,   {Johnstone} R.~M.,  2001, \mn@doi
  [\mnras] {10.1046/j.1365-8711.2001.04243.x}, \href
  {http://adsabs.harvard.edu/abs/2001MNRAS.321L..33F} {321, L33}

\bibitem[\protect\citeauthoryear{{Fabian}, {Sanders}, {Crawford}, {Conselice},
  {Gallagher}  \& {Wyse}}{{Fabian} et~al.}{2003a}]{2003MNRAS.344L..48F}
{Fabian} A.~C.,  {Sanders} J.~S.,  {Crawford} C.~S.,  {Conselice} C.~J.,
  {Gallagher} J.~S.,   {Wyse} R.~F.~G.,  2003a, \mn@doi [\mnras]
  {10.1046/j.1365-8711.2003.06856.x}, \href
  {http://adsabs.harvard.edu/abs/2003MNRAS.344L..48F} {344, L48}

\bibitem[\protect\citeauthoryear{{Fabian}, {Sanders}, {Allen}, {Crawford},
  {Iwasawa}, {Johnstone}, {Schmidt}  \& {Taylor}}{{Fabian}
  et~al.}{2003b}]{2003MNRAS.344L..43F}
{Fabian} A.~C.,  {Sanders} J.~S.,  {Allen} S.~W.,  {Crawford} C.~S.,  {Iwasawa}
  K.,  {Johnstone} R.~M.,  {Schmidt} R.~W.,   {Taylor} G.~B.,  2003b, \mn@doi
  [\mnras] {10.1046/j.1365-8711.2003.06902.x}, \href
  {https://ui.adsabs.harvard.edu/abs/2003MNRAS.344L..43F} {344, L43}

\bibitem[\protect\citeauthoryear{{Fabian}, {Sanders}, {Taylor}, {Allen},
  {Crawford}, {Johnstone}  \& {Iwasawa}}{{Fabian}
  et~al.}{2006}]{2006MNRAS.366..417F}
{Fabian} A.~C.,  {Sanders} J.~S.,  {Taylor} G.~B.,  {Allen} S.~W.,  {Crawford}
  C.~S.,  {Johnstone} R.~M.,   {Iwasawa} K.,  2006, \mn@doi [\mnras]
  {10.1111/j.1365-2966.2005.09896.x}, \href
  {http://adsabs.harvard.edu/abs/2006MNRAS.366..417F} {366, 417}

\bibitem[\protect\citeauthoryear{{Fabian}, {Johnstone}, {Sanders}, {Conselice},
  {Crawford}, {Gallagher}  \& {Zweibel}}{{Fabian}
  et~al.}{2008}]{2008Natur.454..968F}
{Fabian} A.~C.,  {Johnstone} R.~M.,  {Sanders} J.~S.,  {Conselice} C.~J.,
  {Crawford} C.~S.,  {Gallagher} III J.~S.,   {Zweibel} E.,  2008, \mn@doi
  [Nature] {10.1038/nature07169}, \href
  {http://adsabs.harvard.edu/abs/2008Natur.454..968F} {454, 968}

\bibitem[\protect\citeauthoryear{{Fabian}, {Sanders}, {Williams}, {Lazarian},
  {Ferland}  \& {Johnstone}}{{Fabian} et~al.}{2011}]{2011MNRAS.417..172F}
{Fabian} A.~C.,  {Sanders} J.~S.,  {Williams} R.~J.~R.,  {Lazarian} A.,
  {Ferland} G.~J.,   {Johnstone} R.~M.,  2011, \mn@doi [\mnras]
  {10.1111/j.1365-2966.2011.19034.x}, \href
  {http://adsabs.harvard.edu/abs/2011MNRAS.417..172F} {417, 172}

\bibitem[\protect\citeauthoryear{{Fabian} et~al.,}{{Fabian}
  et~al.}{2016}]{2016MNRAS.461..922F}
{Fabian} A.~C.,  et~al., 2016, \mn@doi [\mnras] {10.1093/mnras/stw1350}, \href
  {http://adsabs.harvard.edu/abs/2016MNRAS.461..922F} {461, 922}

\bibitem[\protect\citeauthoryear{{Ferland}, {Fabian}, {Hatch}, {Johnstone},
  {Porter}, {van Hoof}  \& {Williams}}{{Ferland}
  et~al.}{2009}]{2009MNRAS.392.1475F}
{Ferland} G.~J.,  {Fabian} A.~C.,  {Hatch} N.~A.,  {Johnstone} R.~M.,  {Porter}
  R.~L.,  {van Hoof} P.~A.~M.,   {Williams} R.~J.~R.,  2009, \mn@doi [\mnras]
  {10.1111/j.1365-2966.2008.14153.x}, \href
  {http://adsabs.harvard.edu/abs/2009MNRAS.392.1475F} {392, 1475}

\bibitem[\protect\citeauthoryear{{Gendron-Marsolais}
  et~al.,}{{Gendron-Marsolais} et~al.}{2018}]{2018MNRAS.479L..28G}
{Gendron-Marsolais} M.,  et~al., 2018, \mn@doi [\mnras]
  {10.1093/mnrasl/sly084}, \href
  {https://ui.adsabs.harvard.edu/abs/2018MNRAS.479L..28G} {479, L28}

\bibitem[\protect\citeauthoryear{{Hamer} et~al.,}{{Hamer}
  et~al.}{2014}]{2014MNRAS.437..862H}
{Hamer} S.~L.,  et~al., 2014, \mn@doi [\mnras] {10.1093/mnras/stt1949}, \href
  {https://ui.adsabs.harvard.edu/abs/2014MNRAS.437..862H} {437, 862}

\bibitem[\protect\citeauthoryear{{Hamer} et~al.,}{{Hamer}
  et~al.}{2016}]{2016MNRAS.460.1758H}
{Hamer} S.~L.,  et~al., 2016, \mn@doi [\mnras] {10.1093/mnras/stw1054}, \href
  {http://adsabs.harvard.edu/abs/2016MNRAS.460.1758H} {460, 1758}

\bibitem[\protect\citeauthoryear{{Hatch}, {Crawford}, {Fabian}  \&
  {Johnstone}}{{Hatch} et~al.}{2005}]{2005MNRAS.358..765H}
{Hatch} N.~A.,  {Crawford} C.~S.,  {Fabian} A.~C.,   {Johnstone} R.~M.,  2005,
  \mn@doi [\mnras] {10.1111/j.1365-2966.2005.08787.x}, \href
  {https://ui.adsabs.harvard.edu/abs/2005MNRAS.358..765H} {358, 765}

\bibitem[\protect\citeauthoryear{{Heckman}, {Baum}, {van Breugel}  \&
  {McCarthy}}{{Heckman} et~al.}{1989}]{1989ApJ...338...48H}
{Heckman} T.~M.,  {Baum} S.~A.,  {van Breugel} W.~J.~M.,   {McCarthy} P.,
  1989, \mn@doi [\apj] {10.1086/167181}, \href
  {http://adsabs.harvard.edu/abs/1989ApJ...338...48H} {338, 48}

\bibitem[\protect\citeauthoryear{{Hogan}, {McNamara}, {Pulido}, {Nulsen},
  {Russell}, {Vantyghem}, {Edge}  \& {Main}}{{Hogan}
  et~al.}{2017}]{2017ApJ...837...51H}
{Hogan} M.~T.,  {McNamara} B.~R.,  {Pulido} F.,  {Nulsen} P.~E.~J.,  {Russell}
  H.~R.,  {Vantyghem} A.~N.,  {Edge} A.~C.,   {Main} R.~A.,  2017, \mn@doi
  [\apj] {10.3847/1538-4357/aa5f56}, \href
  {https://ui.adsabs.harvard.edu/abs/2017ApJ...837...51H} {837, 51}

\bibitem[\protect\citeauthoryear{{Jaffe}, {Bremer}  \& {Baker}}{{Jaffe}
  et~al.}{2005}]{2005MNRAS.360..748J}
{Jaffe} W.,  {Bremer} M.~N.,   {Baker} K.,  2005, \mn@doi [\mnras]
  {10.1111/j.1365-2966.2005.09073.x}, \href
  {http://adsabs.harvard.edu/abs/2005MNRAS.360..748J} {360, 748}

\bibitem[\protect\citeauthoryear{{Johnstone}, {Fabian}, {Morris}  \&
  {Taylor}}{{Johnstone} et~al.}{2005}]{2005MNRAS.356..237J}
{Johnstone} R.~M.,  {Fabian} A.~C.,  {Morris} R.~G.,   {Taylor} G.~B.,  2005,
  \mn@doi [\mnras] {10.1111/j.1365-2966.2004.08445.x}, \href
  {https://ui.adsabs.harvard.edu/abs/2005MNRAS.356..237J} {356, 237}

\bibitem[\protect\citeauthoryear{{Johnstone}, {Hatch}, {Ferland}, {Fabian},
  {Crawford}  \& {Wilman}}{{Johnstone} et~al.}{2007}]{2007MNRAS.382.1246J}
{Johnstone} R.~M.,  {Hatch} N.~A.,  {Ferland} G.~J.,  {Fabian} A.~C.,
  {Crawford} C.~S.,   {Wilman} R.~J.,  2007, \mn@doi [\mnras]
  {10.1111/j.1365-2966.2007.12460.x}, \href
  {https://ui.adsabs.harvard.edu/abs/2007MNRAS.382.1246J} {382, 1246}

\bibitem[\protect\citeauthoryear{{Kaastra}, {Ferrigno}, {Tamura}, {Paerels},
  {Peterson}  \& {Mittaz}}{{Kaastra} et~al.}{2001}]{2001A&A...365L..99K}
{Kaastra} J.~S.,  {Ferrigno} C.,  {Tamura} T.,  {Paerels} F.~B.~S.,  {Peterson}
  J.~R.,   {Mittaz} J.~P.~D.,  2001, \mn@doi [\aap]
  {10.1051/0004-6361:20000041}, \href
  {http://adsabs.harvard.edu/abs/2001A\%26A...365L..99K} {365, L99}

\bibitem[\protect\citeauthoryear{{Kalberla}, {Burton}, {Hartmann}, {Arnal},
  {Bajaja}, {Morras}  \& {P{\"o}ppel}}{{Kalberla}
  et~al.}{2005}]{2005A&A...440..775K}
{Kalberla} P.~M.~W.,  {Burton} W.~B.,  {Hartmann} D.,  {Arnal} E.~M.,  {Bajaja}
  E.,  {Morras} R.,   {P{\"o}ppel} W.~G.~L.,  2005, \mn@doi [\aap]
  {10.1051/0004-6361:20041864}, \href
  {http://adsabs.harvard.edu/abs/2005A\%26A...440..775K} {440, 775}

\bibitem[\protect\citeauthoryear{{Lakhchaura}, {Saini}  \&
  {Sharma}}{{Lakhchaura} et~al.}{2016}]{2016MNRAS.460.2625L}
{Lakhchaura} K.,  {Saini} T.~D.,   {Sharma} P.,  2016, \mn@doi [\mnras]
  {10.1093/mnras/stw1062}, \href
  {https://ui.adsabs.harvard.edu/abs/2016MNRAS.460.2625L} {460, 2625}

\bibitem[\protect\citeauthoryear{{Li} et~al.,}{{Li}
  et~al.}{2020}]{2020ApJ...889L...1L}
{Li} Y.,  et~al., 2020, \mn@doi [\apjl] {10.3847/2041-8213/ab65c7}, \href
  {https://ui.adsabs.harvard.edu/abs/2020ApJ...889L...1L} {889, L1}

\bibitem[\protect\citeauthoryear{{Liu}, {Pinto}, {Fabian}, {Russell}  \&
  {Sanders}}{{Liu} et~al.}{2019}]{2019MNRAS.485.1757L}
{Liu} H.,  {Pinto} C.,  {Fabian} A.~C.,  {Russell} H.~R.,   {Sanders} J.~S.,
  2019, \mn@doi [\mnras] {10.1093/mnras/stz456}, \href
  {https://ui.adsabs.harvard.edu/abs/2019MNRAS.485.1757L} {485, 1757}

\bibitem[\protect\citeauthoryear{{Lodders} \& {Palme}}{{Lodders} \&
  {Palme}}{2009}]{2009M&PSA..72.5154L}
{Lodders} K.,  {Palme} H.,  2009, Meteoritics and Planetary Science Supplement,
  \href {http://adsabs.harvard.edu/abs/2009M\%26PSA..72.5154L} {72, 5154}

\bibitem[\protect\citeauthoryear{{McDonald}, {Veilleux}, {Rupke}  \&
  {Mushotzky}}{{McDonald} et~al.}{2010}]{2010ApJ...721.1262M}
{McDonald} M.,  {Veilleux} S.,  {Rupke} D. S.~N.,   {Mushotzky} R.,  2010,
  \mn@doi [\apj] {10.1088/0004-637X/721/2/1262}, \href
  {https://ui.adsabs.harvard.edu/abs/2010ApJ...721.1262M} {721, 1262}

\bibitem[\protect\citeauthoryear{{McDonald}, {Veilleux}, {Rupke}, {Mushotzky}
  \& {Reynolds}}{{McDonald} et~al.}{2011}]{2011ApJ...734...95M}
{McDonald} M.,  {Veilleux} S.,  {Rupke} D. S.~N.,  {Mushotzky} R.,   {Reynolds}
  C.,  2011, \mn@doi [\apj] {10.1088/0004-637X/734/2/95}, \href
  {https://ui.adsabs.harvard.edu/abs/2011ApJ...734...95M} {734, 95}

\bibitem[\protect\citeauthoryear{{McDonald}, {Roediger}, {Veilleux}  \&
  {Ehlert}}{{McDonald} et~al.}{2014}]{2014ApJ...791L..30M}
{McDonald} M.,  {Roediger} J.,  {Veilleux} S.,   {Ehlert} S.,  2014, \mn@doi
  [\apjl] {10.1088/2041-8205/791/2/L30}, \href
  {https://ui.adsabs.harvard.edu/abs/2014ApJ...791L..30M} {791, L30}

\bibitem[\protect\citeauthoryear{{McDonald}, {Werner}, {Oonk}  \&
  {Veilleux}}{{McDonald} et~al.}{2015a}]{2015ApJ...804...16M}
{McDonald} M.,  {Werner} N.,  {Oonk} J.~B.~R.,   {Veilleux} S.,  2015a, \mn@doi
  [\apj] {10.1088/0004-637X/804/1/16}, \href
  {https://ui.adsabs.harvard.edu/abs/2015ApJ...804...16M} {804, 16}

\bibitem[\protect\citeauthoryear{{McDonald} et~al.,}{{McDonald}
  et~al.}{2015b}]{2015ApJ...811..111M}
{McDonald} M.,  et~al., 2015b, \mn@doi [\apj] {10.1088/0004-637X/811/2/111},
  \href {https://ui.adsabs.harvard.edu/abs/2015ApJ...811..111M} {811, 111}

\bibitem[\protect\citeauthoryear{{McDonald} et~al.,}{{McDonald}
  et~al.}{2019}]{2019ApJ...885...63M}
{McDonald} M.,  et~al., 2019, \mn@doi [\apj] {10.3847/1538-4357/ab464c}, \href
  {https://ui.adsabs.harvard.edu/abs/2019ApJ...885...63M} {885, 63}

\bibitem[\protect\citeauthoryear{{McNamara} \& {Nulsen}}{{McNamara} \&
  {Nulsen}}{2012}]{2012NJPh...14e5023M}
{McNamara} B.~R.,  {Nulsen} P.~E.~J.,  2012, \mn@doi [New Journal of Physics]
  {10.1088/1367-2630/14/5/055023}, \href
  {https://ui.adsabs.harvard.edu/abs/2012NJPh...14e5023M} {14, 055023}

\bibitem[\protect\citeauthoryear{{McNamara} et~al.,}{{McNamara}
  et~al.}{2014}]{2014ApJ...785...44M}
{McNamara} B.~R.,  et~al., 2014, \mn@doi [\apj] {10.1088/0004-637X/785/1/44},
  \href {https://ui.adsabs.harvard.edu/abs/2014ApJ...785...44M} {785, 44}

\bibitem[\protect\citeauthoryear{{Olivares} et~al.,}{{Olivares}
  et~al.}{2019}]{2019A&A...631A..22O}
{Olivares} V.,  et~al., 2019, \mn@doi [\aap] {10.1051/0004-6361/201935350},
  \href {https://ui.adsabs.harvard.edu/abs/2019A\&A...631A..22O} {631, A22}

\bibitem[\protect\citeauthoryear{{Panagoulia}, {Fabian}  \&
  {Sanders}}{{Panagoulia} et~al.}{2014}]{2014MNRAS.438.2341P}
{Panagoulia} E.~K.,  {Fabian} A.~C.,   {Sanders} J.~S.,  2014, \mn@doi [\mnras]
  {10.1093/mnras/stt2349}, \href
  {https://ui.adsabs.harvard.edu/abs/2014MNRAS.438.2341P} {438, 2341}

\bibitem[\protect\citeauthoryear{{Peres}, {Fabian}, {Edge}, {Allen},
  {Johnstone}  \& {White}}{{Peres} et~al.}{1998}]{1998MNRAS.298..416P}
{Peres} C.~B.,  {Fabian} A.~C.,  {Edge} A.~C.,  {Allen} S.~W.,  {Johnstone}
  R.~M.,   {White} D.~A.,  1998, \mn@doi [\mnras]
  {10.1046/j.1365-8711.1998.01624.x}, \href
  {http://adsabs.harvard.edu/abs/1998MNRAS.298..416P} {298, 416}

\bibitem[\protect\citeauthoryear{{Peterson} et~al.,}{{Peterson}
  et~al.}{2001}]{2001A&A...365L.104P}
{Peterson} J.~R.,  et~al., 2001, \mn@doi [\aap] {10.1051/0004-6361:20000021},
  \href {http://adsabs.harvard.edu/abs/2001A\%26A...365L.104P} {365, L104}

\bibitem[\protect\citeauthoryear{{Piffaretti}, {Arnaud}, {Pratt},
  {Pointecouteau}  \& {Melin}}{{Piffaretti} et~al.}{2011}]{2011A&A...534A.109P}
{Piffaretti} R.,  {Arnaud} M.,  {Pratt} G.~W.,  {Pointecouteau} E.,   {Melin}
  J.~B.,  2011, \mn@doi [\aap] {10.1051/0004-6361/201015377}, \href
  {https://ui.adsabs.harvard.edu/abs/2011A\&A...534A.109P} {534, A109}

\bibitem[\protect\citeauthoryear{{Pinto}, {Kaastra}, {Costantini}  \& {de
  Vries}}{{Pinto} et~al.}{2013}]{2013A&A...551A..25P}
{Pinto} C.,  {Kaastra} J.~S.,  {Costantini} E.,   {de Vries} C.,  2013, \mn@doi
  [\aap] {10.1051/0004-6361/201220481}, \href
  {http://adsabs.harvard.edu/abs/2013A\%26A...551A..25P} {551, A25}

\bibitem[\protect\citeauthoryear{{Pinto} et~al.,}{{Pinto}
  et~al.}{2014}]{2014A&A...572L...8P}
{Pinto} C.,  et~al., 2014, \mn@doi [\aap] {10.1051/0004-6361/201425270}, \href
  {https://ui.adsabs.harvard.edu/abs/2014A\&A...572L...8P} {572, L8}

\bibitem[\protect\citeauthoryear{{Pinto} et~al.,}{{Pinto}
  et~al.}{2015}]{2015A&A...575A..38P}
{Pinto} C.,  et~al., 2015, \mn@doi [\aap] {10.1051/0004-6361/201425278}, \href
  {http://adsabs.harvard.edu/abs/2015A\%26A...575A..38P} {575, A38}

\bibitem[\protect\citeauthoryear{{Pinto} et~al.,}{{Pinto}
  et~al.}{2016}]{2016MNRAS.461.2077P}
{Pinto} C.,  et~al., 2016, \mn@doi [\mnras] {10.1093/mnras/stw1444}, \href
  {http://adsabs.harvard.edu/abs/2016MNRAS.461.2077P} {461, 2077}

\bibitem[\protect\citeauthoryear{{Russell} et~al.,}{{Russell}
  et~al.}{2019}]{2019MNRAS.490.3025R}
{Russell} H.~R.,  et~al., 2019, \mn@doi [\mnras] {10.1093/mnras/stz2719}, \href
  {https://ui.adsabs.harvard.edu/abs/2019MNRAS.490.3025R} {490, 3025}

\bibitem[\protect\citeauthoryear{{Salom{\'e}} \& {Combes}}{{Salom{\'e}} \&
  {Combes}}{2003}]{2003A&A...412..657S}
{Salom{\'e}} P.,  {Combes} F.,  2003, \mn@doi [\aap]
  {10.1051/0004-6361:20031438}, \href
  {https://ui.adsabs.harvard.edu/abs/2003A\&A...412..657S} {412, 657}

\bibitem[\protect\citeauthoryear{{Salom{\'e}} et~al.,}{{Salom{\'e}}
  et~al.}{2006}]{2006A&A...454..437S}
{Salom{\'e}} P.,  et~al., 2006, \mn@doi [\aap] {10.1051/0004-6361:20054745},
  \href {https://ui.adsabs.harvard.edu/abs/2006A\&A...454..437S} {454, 437}

\bibitem[\protect\citeauthoryear{{Salom{\'e}}, {Combes}, {Revaz}, {Downes},
  {Edge}  \& {Fabian}}{{Salom{\'e}} et~al.}{2011}]{2011A&A...531A..85S}
{Salom{\'e}} P.,  {Combes} F.,  {Revaz} Y.,  {Downes} D.,  {Edge} A.~C.,
  {Fabian} A.~C.,  2011, \mn@doi [\aap] {10.1051/0004-6361/200811333}, \href
  {https://ui.adsabs.harvard.edu/abs/2011A\&A...531A..85S} {531, A85}

\bibitem[\protect\citeauthoryear{{Sanders} \& {Fabian}}{{Sanders} \&
  {Fabian}}{2007}]{2007MNRAS.381.1381S}
{Sanders} J.~S.,  {Fabian} A.~C.,  2007, \mn@doi [\mnras]
  {10.1111/j.1365-2966.2007.12347.x}, \href
  {https://ui.adsabs.harvard.edu/abs/2007MNRAS.381.1381S} {381, 1381}

\bibitem[\protect\citeauthoryear{{Sanders} \& {Fabian}}{{Sanders} \&
  {Fabian}}{2011}]{2011MNRAS.412L..35S}
{Sanders} J.~S.,  {Fabian} A.~C.,  2011, \mn@doi [\mnras]
  {10.1111/j.1745-3933.2010.01000.x}, \href
  {http://adsabs.harvard.edu/abs/2011MNRAS.412L..35S} {412, L35}

\bibitem[\protect\citeauthoryear{{Sanders}, {Fabian}, {Allen}, {Morris},
  {Graham}  \& {Johnstone}}{{Sanders} et~al.}{2008}]{2008MNRAS.385.1186S}
{Sanders} J.~S.,  {Fabian} A.~C.,  {Allen} S.~W.,  {Morris} R.~G.,  {Graham}
  J.,   {Johnstone} R.~M.,  2008, \mn@doi [\mnras]
  {10.1111/j.1365-2966.2008.12952.x}, \href
  {http://adsabs.harvard.edu/abs/2008MNRAS.385.1186S} {385, 1186}

\bibitem[\protect\citeauthoryear{{Sanders}, {Fabian}  \& {Taylor}}{{Sanders}
  et~al.}{2009}]{2009MNRAS.396.1449S}
{Sanders} J.~S.,  {Fabian} A.~C.,   {Taylor} G.~B.,  2009, \mn@doi [\mnras]
  {10.1111/j.1365-2966.2009.14892.x}, \href
  {https://ui.adsabs.harvard.edu/abs/2009MNRAS.396.1449S} {396, 1449}

\bibitem[\protect\citeauthoryear{{Sanders}, {Fabian}, {Frank}, {Peterson}  \&
  {Russell}}{{Sanders} et~al.}{2010}]{2010MNRAS.402..127S}
{Sanders} J.~S.,  {Fabian} A.~C.,  {Frank} K.~A.,  {Peterson} J.~R.,
  {Russell} H.~R.,  2010, \mn@doi [\mnras] {10.1111/j.1365-2966.2009.15902.x},
  \href {https://ui.adsabs.harvard.edu/abs/2010MNRAS.402..127S} {402, 127}

\bibitem[\protect\citeauthoryear{{Sanders} et~al.,}{{Sanders}
  et~al.}{2016}]{2016MNRAS.457...82S}
{Sanders} J.~S.,  et~al., 2016, \mn@doi [\mnras] {10.1093/mnras/stv2972}, \href
  {https://ui.adsabs.harvard.edu/abs/2016MNRAS.457...82S} {457, 82}

\bibitem[\protect\citeauthoryear{{Simionescu}, {Tremblay}, {Werner}, {Canning},
  {Allen}  \& {Oonk}}{{Simionescu} et~al.}{2018}]{2018MNRAS.475.3004S}
{Simionescu} A.,  {Tremblay} G.,  {Werner} N.,  {Canning} R.~E.~A.,  {Allen}
  S.~W.,   {Oonk} J.~B.~R.,  2018, \mn@doi [\mnras] {10.1093/mnras/sty047},
  \href {https://ui.adsabs.harvard.edu/abs/2018MNRAS.475.3004S} {475, 3004}

\bibitem[\protect\citeauthoryear{{Sparks}, {Macchetto}  \& {Golombek}}{{Sparks}
  et~al.}{1989}]{1989ApJ...345..153S}
{Sparks} W.~B.,  {Macchetto} F.,   {Golombek} D.,  1989, \mn@doi [\apj]
  {10.1086/167890}, \href
  {https://ui.adsabs.harvard.edu/abs/1989ApJ...345..153S} {345, 153}

\bibitem[\protect\citeauthoryear{{Sparks}, {Ford}  \& {Kinney}}{{Sparks}
  et~al.}{1993}]{1993ApJ...413..531S}
{Sparks} W.~B.,  {Ford} H.~C.,   {Kinney} A.~L.,  1993, \mn@doi [\apj]
  {10.1086/173022}, \href
  {https://ui.adsabs.harvard.edu/abs/1993ApJ...413..531S} {413, 531}

\bibitem[\protect\citeauthoryear{{Tamura} et~al.,}{{Tamura}
  et~al.}{2001}]{2001A&A...365L..87T}
{Tamura} T.,  et~al., 2001, \mn@doi [\aap] {10.1051/0004-6361:20000038}, \href
  {http://adsabs.harvard.edu/abs/2001A\%26A...365L..87T} {365, L87}

\bibitem[\protect\citeauthoryear{{Taylor}, {Fabian}, {Gentile}, {Allen},
  {Crawford}  \& {Sanders}}{{Taylor} et~al.}{2007}]{2007MNRAS.382...67T}
{Taylor} G.~B.,  {Fabian} A.~C.,  {Gentile} G.,  {Allen} S.~W.,  {Crawford} C.,
    {Sanders} J.~S.,  2007, \mn@doi [\mnras]
  {10.1111/j.1365-2966.2007.12368.x}, \href
  {https://ui.adsabs.harvard.edu/abs/2007MNRAS.382...67T} {382, 67}

\bibitem[\protect\citeauthoryear{{Vantyghem} et~al.,}{{Vantyghem}
  et~al.}{2016}]{2016ApJ...832..148V}
{Vantyghem} A.~N.,  et~al., 2016, \mn@doi [\apj] {10.3847/0004-637X/832/2/148},
  \href {https://ui.adsabs.harvard.edu/abs/2016ApJ...832..148V} {832, 148}

\bibitem[\protect\citeauthoryear{{Vantyghem} et~al.,}{{Vantyghem}
  et~al.}{2017}]{2017ApJ...848..101V}
{Vantyghem} A.~N.,  et~al., 2017, \mn@doi [\apj] {10.3847/1538-4357/aa8fd0},
  \href {https://ui.adsabs.harvard.edu/abs/2017ApJ...848..101V} {848, 101}

\bibitem[\protect\citeauthoryear{{Walker}, {Kosec}, {Fabian}  \&
  {Sanders}}{{Walker} et~al.}{2015}]{2015MNRAS.453.2480W}
{Walker} S.~A.,  {Kosec} P.,  {Fabian} A.~C.,   {Sanders} J.~S.,  2015, \mn@doi
  [\mnras] {10.1093/mnras/stv1829}, \href
  {https://ui.adsabs.harvard.edu/abs/2015MNRAS.453.2480W} {453, 2480}

\bibitem[\protect\citeauthoryear{{Werner} et~al.,}{{Werner}
  et~al.}{2010}]{2010MNRAS.407.2063W}
{Werner} N.,  et~al., 2010, \mn@doi [\mnras]
  {10.1111/j.1365-2966.2010.16755.x}, \href
  {http://adsabs.harvard.edu/abs/2010MNRAS.407.2063W} {407, 2063}

\bibitem[\protect\citeauthoryear{{Werner} et~al.,}{{Werner}
  et~al.}{2011}]{2011MNRAS.415.3369W}
{Werner} N.,  et~al., 2011, \mn@doi [\mnras]
  {10.1111/j.1365-2966.2011.18957.x}, \href
  {https://ui.adsabs.harvard.edu/abs/2011MNRAS.415.3369W} {415, 3369}

\bibitem[\protect\citeauthoryear{{Werner} et~al.,}{{Werner}
  et~al.}{2013}]{2013ApJ...767..153W}
{Werner} N.,  et~al., 2013, \mn@doi [\apj] {10.1088/0004-637X/767/2/153}, \href
  {https://ui.adsabs.harvard.edu/abs/2013ApJ...767..153W} {767, 153}

\bibitem[\protect\citeauthoryear{{White}, {Jones}  \& {Forman}}{{White}
  et~al.}{1997}]{1997MNRAS.292..419W}
{White} D.~A.,  {Jones} C.,   {Forman} W.,  1997, \mn@doi [\mnras]
  {10.1093/mnras/292.2.419}, \href
  {http://adsabs.harvard.edu/abs/1997MNRAS.292..419W} {292, 419}

\bibitem[\protect\citeauthoryear{{Willingale}, {Starling}, {Beardmore},
  {Tanvir}  \& {O'Brien}}{{Willingale} et~al.}{2013}]{2013MNRAS.431..394W}
{Willingale} R.,  {Starling} R.~L.~C.,  {Beardmore} A.~P.,  {Tanvir} N.~R.,
  {O'Brien} P.~T.,  2013, \mn@doi [\mnras] {10.1093/mnras/stt175}, \href
  {http://adsabs.harvard.edu/abs/2013MNRAS.431..394W} {431, 394}

\bibitem[\protect\citeauthoryear{{Wilman}, {Edge}, {Johnstone}, {Fabian},
  {Allen}  \& {Crawford}}{{Wilman} et~al.}{2002}]{2002MNRAS.337...63W}
{Wilman} R.~J.,  {Edge} A.~C.,  {Johnstone} R.~M.,  {Fabian} A.~C.,  {Allen}
  S.~W.,   {Crawford} C.~S.,  2002, \mn@doi [\mnras]
  {10.1046/j.1365-8711.2002.05791.x}, \href
  {https://ui.adsabs.harvard.edu/abs/2002MNRAS.337...63W} {337, 63}

\bibitem[\protect\citeauthoryear{{Wilman}, {Edge}  \& {Johnstone}}{{Wilman}
  et~al.}{2005}]{2005MNRAS.359..755W}
{Wilman} R.~J.,  {Edge} A.~C.,   {Johnstone} R.~M.,  2005, \mn@doi [\mnras]
  {10.1111/j.1365-2966.2005.08956.x}, \href
  {https://ui.adsabs.harvard.edu/abs/2005MNRAS.359..755W} {359, 755}

\bibitem[\protect\citeauthoryear{{de Plaa}, {Kaastra}, {Werner}, {Pinto}  \&
  {Kosec}}{{de Plaa} et~al.}{2017}]{2017A&A...607A..98D}
{de Plaa} J.,  {Kaastra} J.~S.,  {Werner} N.,  {Pinto} C.,   {Kosec} P. e.~a.,
  2017, \mn@doi [A\&A] {10.1051/0004-6361/201629926}, \href
  {http://adsabs.harvard.edu/abs/2017A\%26A...607A..98D} {607, A98}

\makeatother
\end{thebibliography}



\appendix
\section{}

We summarise the previous spatial analyses of the inner core which involved different gas phases studied in this work.

\noindent \textbf{2A0335+096}\, The optical image shows a 20 kpc bar feature (\citealt{2007AJ....134...14D}), 
which is surrounded by the soft X-ray emitting gas below 1 keV (\citealt{2009MNRAS.396.1449S}). 
The morphology of the molecular filament matches well to the peak of the optical emission (\citealt{2016ApJ...832..148V}).

\noindent \textbf{A262}\, The molecular gas is in the form of a compact disk around the nucleus (\citealt{2019MNRAS.490.3025R}; \citealt{2019A&A...631A..22O}). 
It spatially coincides with the optical image and the soft X-ray gas at  $\sim$0.5 keV (\citealt{2010MNRAS.402..127S}). 
All these components are surrounded by the soft X-ray gas at $\sim$0.9 keV.

\noindent \textbf{A496}\, The filaments in the H$\alpha$ nebula do not extend beyond the presence of the X-ray cold front at 15 kpc (\citealt{2010ApJ...721.1262M}). 

\noindent \textbf{A2052}\, Two H$\alpha$ emitting regions are separated by the X-ray cavity (\citealt{2011ApJ...734...95M}; \citealt{2011ApJ...737...99B}).
The southern nebula covers the X-ray brightest cluster centre and two filaments extend along the soft X-ray emitting region.
The northern H$\alpha$ emitting region is located 8-15 kpc from the centre. 
This morphology is the same as the soft X-ray emitting gas at $\sim$0.8 keV gas. 

\noindent \textbf{A3526} The core of the Centaurus cluster has a soft X-ray plume structure about 10 kpc across (\citealt{2016MNRAS.457...82S}). 
The same structure has also been observed in the optical and radio band (e.g. \citealt{1989ApJ...345..153S}; \citealt{2005MNRAS.363..216C}; \citealt{2007MNRAS.382...67T}; \citealt{2016MNRAS.461..922F}). 
The molecular gas is in clumps and filaments within the H$\alpha$ nebula (\citealt{2019A&A...631A..22O}).

\noindent \textbf{A3581} The thin H$\alpha$ filaments extend $\sim$ 13 kpc (30 arcsec) from the centre on both sides and surround the molecular gas which is less than 5 kpc across (\citealt{2019A&A...631A..22O}). 
They align with the X-ray emitting gas below 1 keV (\citealt{2013MNRAS.435.1108C}). 

\noindent \textbf{AS1101} Also known as S\'ersic 159-03, the cluster has a $\sim$ 40 kpc long H$\alpha$ filament which extends along the far UV emission (\citealt{2011MNRAS.415.3369W}; \citealt{2015ApJ...804...16M}). 
Most molecular gas is in small clumps of a few kpc within the H$\alpha$ filaments (\citealt{2019A&A...631A..22O}).

\noindent \textbf{Perseus} The H$\alpha$ nebula in the Perseus cluster is very extended (\citealt{2001AJ....122.2281C}; \citealt{2008Natur.454..968F}; \citealt{2018MNRAS.479L..28G}), and coincides with the distributing molecular gas (\citealt{2006A&A...454..437S,2011A&A...531A..85S}). 
These also correspond with the 0.5-1 keV gas (\citealt{2007MNRAS.381.1381S}). 
It is found that the X-ray cooling flow down to 0.25 keV can produce the same flux as the H$\alpha$ in filaments. 
\citet{2015MNRAS.453.2480W} also found that the X-ray gas surrounding the H$\alpha$ filaments can be modelled by a charge exchange component in addition to the standard thermal component.
This provides a potential pathway for the energy transport between the hot ICM and the H$\alpha$ filaments.

\noindent \textbf{Virgo} Two major filaments of H$\alpha$ and 0.7-0.9 keV X-ray emitting gas are found in the Virgo cluster, both of which are less than 2 kpc long (\citealt{2010MNRAS.407.2063W,2013ApJ...767..153W}). 
The molecular filament is very small (0.6 kpc) and less massive than any other clusters in our sample (\citealt{2018MNRAS.475.3004S}; \citealt{2019A&A...631A..22O}). 

\bsp	
\label{lastpage}
\end{document}